\documentclass[10pt, conference, letterpaper]{IEEEtran}
\IEEEoverridecommandlockouts
\usepackage{cite}
\usepackage{amsmath,amssymb,amsfonts}
\usepackage{algorithmic}
\usepackage{graphicx}
\usepackage{textcomp}
\usepackage{xcolor}
\usepackage{comment}
\usepackage{float}
\usepackage{url}
\usepackage{caption}
\usepackage{subfigure}

\newtheorem{proposition}{Proposition}

\def\BibTeX{{\rm B\kern-.05em{\sc i\kern-.025em b}\kern-.08em
    T\kern-.1667em\lower.7ex\hbox{E}\kern-.125emX}}
\begin{document}

\title{Towards the Tradeoff Between Service Performance and Information Freshness}

\author{Zhongdong Liu and Bo Ji
\thanks{This work was supported in part by the NSF under Grants CCF-1657162 and CNS-1651947.}
\thanks{Zhongdong Liu (zhongdong.liu@temple.edu) and Bo Ji (boji@temple.edu) are with the Department of Computer and Information Sciences, Temple University, Philadelphia, PA.}
}

\maketitle

\begin{abstract}
The last decade has witnessed an unprecedented growth in the demand for data-driven real-time services. These services are fueled by emerging applications that require rapidly injecting data streams and computing updated analytics results in real-time (or near-real-time). In many of such applications, the computing resources are often shared for processing both updates from information sources and queries from end users. This requires joint scheduling of updates and queries because the service provider needs to make a critical decision upon receiving a user query: either it responds immediately with currently available but possibly stale information, or it first processes new updates and then responds with fresher information. Hence, the tradeoff between \emph{service performance} (e.g., response time) and \emph{information freshness} naturally arises in this context. To that end, we propose a simple \emph{single-server two-queue model} that captures the coupled scheduling of updates and queries and aim to design scheduling policies that can properly address the important tradeoff between performance and freshness. Specifically, we consider the \emph{response time} as a performance metric and the \emph{Age of Information (AoI)} as a freshness metric. After demonstrating the limitations of the simplest First-Come-First-Served (FCFS) policy, we propose two \emph{threshold-based policies}: the Query-$k$ policy that prioritizes queries and the Update-$k$ policy that prioritizes updates. Then, we rigorously analyze both the response time and the \emph{Peak AoI (PAoI)} of the threshold-based policies.
Further, we propose the Joint-$(M,N)$ policy, which allows flexibly prioritizing updates or queries through choosing different values of two thresholds $M$ and $N$.
Finally, we conduct simulations to evaluate the response time and the PAoI of the proposed policies. The results show that our proposed threshold-based policies can effectively control the balance between performance and freshness.

\end{abstract}


\section{Introduction}

The last decade has witnessed an unprecedented growth in the demand for data-driven real-time services
(built on frameworks such as Apache Storm~\cite{storm}).
These services are fueled by emerging applications that require rapidly injecting data streams and computing updated analytics results in \emph{real-time} (or \emph{near-real-time}). 
For such applications, \emph{service performance} (e.g., response time) perceived by end users is typically a primary concern and has been extensively studied in the literature.
Yet, \emph{freshness} of the information received by end users, another equally or even more important concern, has not received enough attention.
Unilaterally optimizing service performance without accounting for information freshness could render users receive stale information, which is potentially of much less value or even useless.
For example, upon receiving a user query, in order to minimize the response time the service provider may respond immediately with currently available but possibly outdated information. On the other hand, it may choose to first process new updates and then responds with fresher information if the goal is to optimize freshness.
Hence, \emph{there exists a natural tradeoff between service performance (e.g., response time) and information freshness}. 


In this paper, we consider the \emph{response time} as a performance metric and the \emph{Age of Information (AoI)}~\cite{kaul2012real} as a freshness metric. While the response time has been shared as a standard performance metric, the AoI, which is defined as the time elapsed since the generation of the freshest update among those that have been delivered to the receiver (see Section~\ref{sec:model} for the formal definition), is a recently proposed freshness/timeliness metric~\cite{kaul2012real}. 
Note that there is a limited body of existing work (see, e.g., \cite{labrinidis2004exploring,qu2006unit,qu2007preference}) that investigates the important tradeoff between performance and freshness as we do. However, all of these studies provide heuristic solutions only and fall short of theoretical results with rigorous analysis. 

To that end, in this paper we aim to fill this important gap and design efficient policies that can properly address the critical tradeoff between performance and freshness. We summarize the main contributions of this paper as follows.

First, we propose a simple \emph{single-server two-queue model} that captures the coupled scheduling of updates and queries. 
Second, after demonstrating the limitations of the First-Come-First-Served (FCFS) policy, we propose two \emph{threshold-based policies}: the Query-$k$ policy that prioritizes queries and the Update-$k$ policy that prioritizes updates. Then, we rigorously analyze the response time and the Peak AoI (PAoI) (i.e., the maximum value of the AoI at the server immediately before a new update is processed)~\cite{costa2014age} of these two policies.
To the best of knowledge, \emph{this is the first analytical work that systematically studies the tradeoff between performance and freshness in a rigorous manner}.
Further, we propose the Joint-$(M,N)$ policy, which allows flexibly prioritizing updates or queries through choosing different values of two thresholds $M$ and $N$.
Finally, we conduct simulations to evaluate the response time and the PAoI of the proposed policies. The results show that our proposed threshold-based policies can effectively control the balance between performance and freshness.

The rest of this paper is organized as follows. We first discuss related work in Section~\ref{sec:relatedwork}. Then, we describe our proposed model in Section~\ref{sec:model}. In Section~\ref{sec:analyses}, we analyze the response time and the PAoI of our proposed threshold-based policies, followed by a discussion on the simulation results in Section~\ref{sec:simulation}. Finally, we make concluding remarks in Section~\ref{sec:conclusion}.

\section{Related Work} \label{sec:relatedwork}
Research on the performance (e.g., response time) started very early. In \cite{harchol1997queue}, it studies under what condition such that  the performance in FCFS policy is better than that in \emph{Process-Sharing (PS)} policy. The authors conclude that a special task assignment which has the ability to inspect incoming tasks and assign them to hosts for service can achieve this goal.  Further, the performance comparison  between the  \emph{Shortest-Remaining-Processing-Time (SRPT)} policy and the PS policy is studied in \cite{harchol1998case}, it shows SRPT  has better performance than PS when the service load is high. Based on the SRPT policy, the work of \cite{harchol2003size} improves the performance by giving preference to the queries whose remaining size or original size is small. The simulation results show that even the queries for large files suffer little in this SRPT-based scheduling. The first analytical study of performance and robustness in threshold-based resource allocation policies appears in \cite{osogami2005robustness}, where the authors conclude that using multiple thresholds does not always provide benefits to the performance and robustness. However, there is still no analytical work considering the freshness in these studies. 

The notion of AoI is formally introduced in \cite{kaul2012real}, where the authors analyze the time average AoI in M/M/1, M/D/1, and D/M/1 systems under the FCFS policy. Since this seminal work, the study on the AoI has attracted a lot of research interests. 
There is a large body of work that focuses on the analysis of the AoI under a number of queueing model. For example, the work of \cite{costa2014age,kam2013age,yates2012real} focuses on the model where the updates arrive according to the Poisson process and are served by a single server.  
There is another body of work that considers how to minimize the AoI by carefully designing scheduling policies in different scenarios (e.g., wireless networks~\cite{kadota2018scheduling,lu2018age} and energy harvesting networks~\cite{yates2015lazy,sun2017update}).
In \cite{sang2017power}, the authors propose the Pull model for investigating the expected AoI at the user's side and discover a new tradeoff between different levels of information freshness and different response times across the servers.
Besides the above work that focuses on the analysis and optimization of the AoI, several other work also considers applications where the AoI is highly relevant (see, e.g., \cite{patra2016minimizing,he2014practical}). 

Despite the aforementioned studies on service performance and information freshness, the tradeoff between them has often been neglected in the literature (partially due to the nature of the considered applications), except for the following limited work. In \cite{labrinidis2004exploring}, 
the tradeoff of performance and freshness has been considered for database-driven web servers, where the goal is to optimize performance under the freshness constraint. The work of \cite{qu2006unit} proposes to combine performance and freshness into a single compound metric and addresses the tradeoff between them through optimizing the compound metric. Further, the work of \cite{qu2006unit} has been extended to account for user preference for performance and freshness~\cite{qu2007preference}. In stark contrast to these studies that provide heuristic solutions only, in this paper we aim to systematically understand this tradeoff by providing theoretical results with rigorous analysis.

\section{System Model} \label{sec:model}

In this section, we describe the single-server two-queue model and give the formal definition of the AoI and the PAoI. 

We consider a queueing system where a single server is shared by two M/M/1 queues. One is the update queue that buffers updates coming from the information source, and the other is the query queue that buffers queries from the user. We assume that the arrival processes of the updates and the queries are both Poisson with rate ${\lambda_u}$ and ${\lambda_q}$, respectively. Also, we assume that the service times of the updates and the queries are both exponentially distributed with mean $1/\mu_u$ and $1/\mu_q$, respectively. Therefore, the loads of the update queue and query queue can be denoted by ${\rho _u} = {\lambda _u}/{\mu _u}$ and ${\rho _q} = {\lambda _q}/{\mu _q}$, respectively.
In addition, we assume that the server does not remain at an empty queue if the other queue is nonempty. Further, let $X_{u,i}$ be the inter-arrival time between the $i$-th update and the $(i-1)$-th update, let $S_{u,i}$ be the service time of the $i$-th update, let $T_{u,i}$ be the system time of the $i$-th update, and let $N_{u,i}^u$ (resp., $N_{u,i}^q$) be the number of updates (resp., queries) seen by the $i$-th update upon its arrival. More generally, we drop subscript $i$ and use $X_{u}$, $S_{u}$, and $T_{u}$ to denote the corresponding quantities for an ordinary update. For example, $X_u$ denotes the inter-arrival time of an update.
Similarly, we define $X_{q,i}$, $S_{q,i}$, $T_{q,i}$, $N_{q,i}^u$, $N_{q,i}^q$,  $X_{q}$, $S_{q}$, and $T_{q}$ for queries.  Also, we use $N_u$ (resp., $N_q$) to denote the number of updates (resp., queries) in the system. 

\begin{figure}[!t]
	\centering
	\includegraphics[scale=0.5]{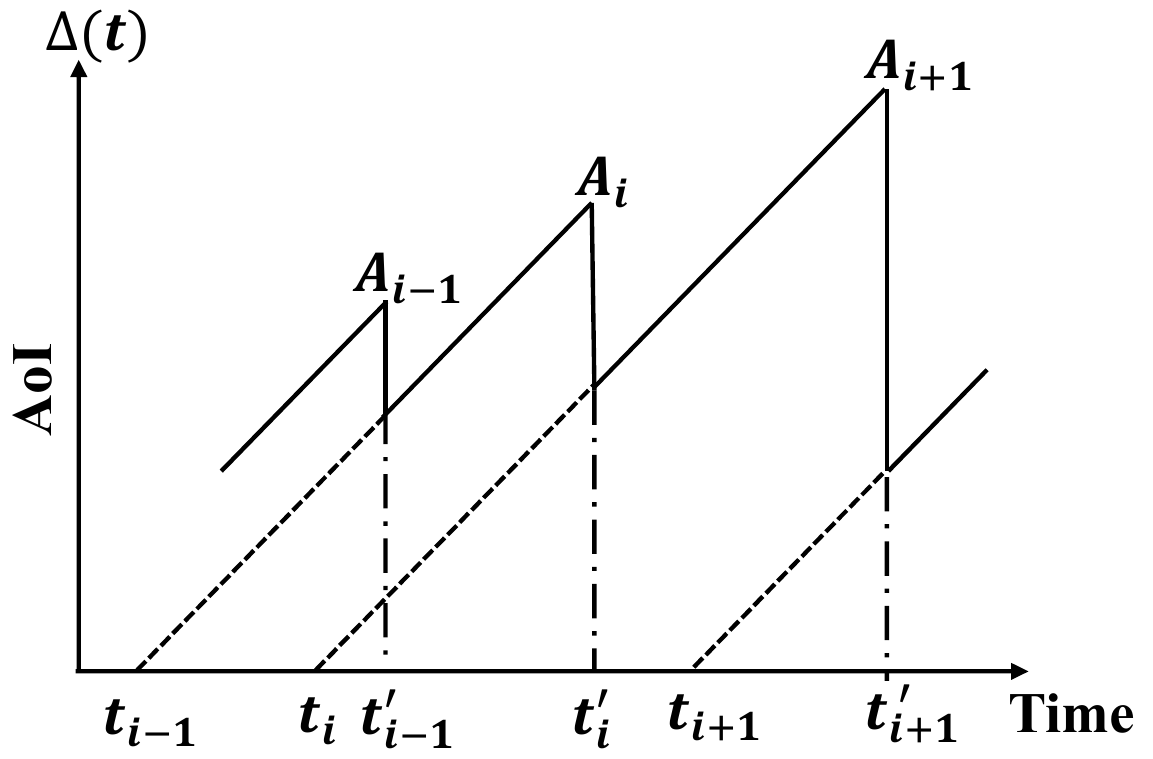}
	\caption{An example of the AoI evolution}
	\label{fig:aoi_ex}
\end{figure}

Next, we give the formal definition of the AoI and the PAoI. Let $U\left( t \right)$ denote the generation time of the freshest update among those that have been processed by the server. We use $\Delta(t)$ to denote the AoI at time $t$, which is defined as the time elapsed since the generation of this freshest update, i.e., $\Delta(t) \triangleq t - U(t)$. An example of the AoI evolution is shown in  Fig.~\ref{fig:aoi_ex}. The AoI increases linearly as time goes until a new update is completely processed. For example, consider the $i$-th update, which is generated at time $t_i$ and finishes processing at time $t^{\prime}_i$. When the server finishes processing the $i$-th update, the AoI drops to the value of $t^{\prime}_i-t_i$, i.e., the system time of the $i$-th update. Then, the average AoI can be defined as 
\begin{equation}
    \Delta  = \mathop {\lim }\limits_{\tau  \to \infty } \dfrac{1}{\tau }\int_0^\tau  {\Delta \left( t \right)} dt.
\end{equation}

Analyzing the average AoI involves two important quantities: the inter-arrival time and the system time of the updates. The fact that the latter is dependent on the former often renders the analysis of the average AoI quite challenging except for some simple settings (e.g., M/M/1 queue)~\cite{kaul2012real}. 
On the other hand, the analysis of the average PAoI is usually more tractable. The PAoI is the maximum value of the AoI achieved immediately before a new update is processed. Let $A_i$ be the $i$-th PAoI. From Fig.~\ref{fig:aoi_ex}, we can see $A_i = t^{\prime}_i - t_{i - 1}$. This can be rewritten as the sum of the inter-arrival time between the $i$-th update and the previous update (i.e., $t_i - t_{i - 1}$) and the system time of the $i$-th update (i.e., $t^{\prime}_i - {t_i}$). Therefore, the expected PAoI can be expressed as
\begin{equation}
\mathbb{E}\left[ A \right] = \mathbb{E}\left[ X_u \right] + \mathbb{E}\left[ T_u \right],
\label{eq:paoi}
\end{equation} 
where $\mathbb{E}\left[  \cdot  \right]$ is the expectation operator and $A$ is the PAoI corresponding to an update. While computing the first term of the right hand side (RHS) of Eq.~\eqref{eq:paoi} is trivial, i.e., $\mathbb{E}\left[ X_u \right] = 1/\lambda_u$, computing the second term $\mathbb{E}\left[ T_u \right]$ is more involved as it depends on the underlying scheduling policy. 

To measure the service performance, we consider the average response time, i.e., the system time of the queries $T_{q}$.

For quick reference, we provide a summary of the key notations of this paper in Table~\ref{tab:notation}.

\begin{table}[!t]
\centering
\caption{Summary of Key Notations}
\begin{tabular}{l|l}
\hline
Symbol & Meaning \\  \hline
${\lambda _u}$      &     Arrival rate of the updates  \\
${\mu _u}$    &    Service rate of the updates    \\  
${\rho _u}$   &     Load of the update queue (i.e., ${\rho _u} = {\lambda _u}/{\mu _u}$)      \\
$X_{u,i}$  &    Inter-arrival time between the $i$-th and $(i-1)$-th updates   \\
$S_{u,i}$  &    Service time of  the $i$-th update  \\
$T_{u,i}$  &    System time of the  $i$-th update   \\
$N_{u,i}^u$  &   Number of updates seen by the $i$-th update upon arrival   \\
$N_{u,i}^q$  &   Number of queries seen by the $i$-th update upon arrival   \\
$N_u$  &   Number of updates in the system   \\
$A_i$  &    The $i$-th PAoI   \\

${\lambda _q}$    &    Arrival rate of the queries     \\
${\mu _q}$    &    Service rate of the queries    \\
${\rho _q}$       &    Load of the query queue (i.e., ${\rho _q} = {\lambda _q}/{\mu _q}$)       \\
$X_{q,i}$  &    Inter-arrival time between the $i$-th and $(i-1)$-th queries   \\
$S_{q,i}$  &    Service time of  the $i$-th query   \\
$T_{q,i}$  &    System time (or response time) of the  $i$-th query    \\
$N_{q,i}^u$  &   Number of updates seen by the $i$-th query upon arrival  \\
$N_{q,i}^q$  &   Number of queries seen by the $i$-th query upon arrival  \\
$N_q$  &   Number of queries in the system   \\
$\rho $  &    Total load (i.e., $\rho  = {\rho _u} + {\rho _q}$)  \\
\hline         
\end{tabular}
\label{tab:notation}
\end{table}

\section{Scheduling Policies} \label{sec:analyses}
In this section, we first consider a simple scheduling policy, the FCFS policy, and explain its limitation in balancing the service performance and information freshness. Then, we propose two threshold-based policies: the Query-$k$ policy that prioritizes queries and the Update-$k$ policy that prioritizes updates, and rigorously analyze the response time and the PAoI under these policies. Further, we propose the Joint-$(M,N)$ policy, where we jointly set thresholds $M$ and $N$ for the updates and the queries, respectively. The Joint-$(M,N)$ policy generalizes the Query-$k$ policy and the Update-$k$ policy and allows flexibly prioritizing updates or queries through choosing different values of $M$ and $N$.

\subsection{The FCFS Policy}
We first consider the FCFS policy, a simple policy that serves updates and queries according to the order of their arrivals. Preemption is not allowed during the service. The main results for the FCFS policy are stated in Proposition~\ref{pro:fcfs}. 

\begin{proposition}
Under the FCFS policy, the expected response time is 
\begin{equation} \label{eq:fcfs_resptime}
\mathbb{E}\left[ {{T_q}} \right] = \dfrac{{{\rho _u}/{\mu _u} + \left( {1 - {\rho _u}} \right)/{\mu _q}}}{{1 - {\rho _u} - {\rho _q}}},
\end{equation}
and the expected PAoI is
\begin{equation} \label{eq:fcfs_paoi}
\mathbb{E}\left[ {{A}} \right] = \dfrac{1}{{{\lambda _u}}} + \dfrac{{{\rho _q}/{\mu _q} + \left( {1 - {\rho _q}} \right)/{\mu _u}}}{{1 - {\rho _u} - {\rho _q}}}.
\end{equation}
\label{pro:fcfs}
\end{proposition}

\begin{IEEEproof}
Consider an arbitrary update, say the $i$-th update. It is easy to see that its system time consists of the following three components: (\romannumeral1) the total service time of all the other updates that are already in the system upon its arrival; (\romannumeral2) the total service time of all the queries are already in the system upon upon its arrival; (\romannumeral3) its own service time.
Taking the sum of these three components, we have 
\begin{equation}  \label{ithUpdateSystemTime}
{T_{u,i}} = \sum\limits_{j = i_u^*}^{i_u^* + N_{u,i}^u - 1} {{S_{u,j}}}  + \sum\limits_{k = i_q^*}^{i_q^* + N_{u,i}^q - 1} {{S_{q,k}}}  + {S_{u,i}},
\end{equation}
where ${i_u^*}$ (resp., ${i_q^*}$) is the index of first update (resp., query) seen by the $i$-th update upon its arrival to the system. 

Taking the expectation of both sides of Eq.~(\ref{ithUpdateSystemTime}), we have 
\begin{equation} \label{expectationOfithUpdateSystemTime}
    \mathbb{E}\left[ {{T_{u,i}}} \right] = \mathbb{E}\left[ {\sum\limits_{j = i_u^*}^{i_u^* + N_{u,i}^u - 1} {{S_{u,j}}} } \right] + \mathbb{E}\left[ {\sum\limits_{k = i_q^*}^{i_q^* + N_{u,i}^q - 1} {{S_{q,k}}} } \right] + \mathbb{E}\left[ {{S_{u,i}}} \right].
\end{equation}
Note that in Eq.~\eqref{expectationOfithUpdateSystemTime}, the ${{S_{u,j}}}$'s are \emph{i.i.d.} with mean $\mathbb{E}\left[ {{S_u}} \right]$, and they are independent of $N_{u,i}^u$. Similarly, the ${{S_{q,k}}}$'s are \emph{i.i.d.} with mean $\mathbb{E}\left[ {{S_q}} \right]$, and they are independent of $N_{u,i}^q$. Hence, by applying Wald's equation to the first term and the second term of the RHS of Eq.~\eqref{expectationOfithUpdateSystemTime}, respectively, we have
\begin{equation}  \label{expectationOfithUpdateWaitingServiceTime}
\mathbb{E}\left[ {\sum\limits_{k = i_q^*}^{i_q^* + N_{u,i}^q - 1} {{S_{q,k}}} } \right] = \mathbb{E}\left[ {{S_q}} \right] \cdot \mathbb{E}\left[ {N_q} \right]
\end{equation}
and
\begin{equation}  \label{expectationOfithQueryWaitingServiceTime}
\mathbb{E}\left[ {\sum\limits_{j = i_u^*}^{i_u^* + N_{u,i}^u - 1} {{S_{u,j}}} } \right] = \mathbb{E}\left[ {{S_u}} \right] \cdot \mathbb{E}\left[ {N_u} \right].
\end{equation}
Then, by plugging Eqs.~\eqref{expectationOfithUpdateWaitingServiceTime} and \eqref{expectationOfithQueryWaitingServiceTime} into Eq.~\eqref{expectationOfithUpdateSystemTime} and using the fact that $\mathbb{E}\left[ {{S_{u,i}}} \right] = \mathbb{E}\left[ {{S_u}} \right]$,  we have
\begin{equation} \label{avgSystemTimeOfUpdates}
    \mathbb{E}\left[ {{T_u}} \right] = \mathbb{E}\left[ {{N_u}} \right] \cdot \mathbb{E}\left[ {{S_u}} \right] + \mathbb{E}\left[ {{N_q}} \right] \cdot \mathbb{E}\left[ {{S_q}} \right] + \mathbb{E}\left[ {{S_u}} \right].
\end{equation}
By applying the Little's Law (i.e., $\mathbb{\mathbb{E}}\left[ {N_u} \right] = \lambda_u \mathbb{\mathbb{E}}\left[ {T_u} \right]$ and $\mathbb{\mathbb{E}}\left[ {N_q} \right] = \lambda_q \mathbb{\mathbb{E}}\left[ {T_q} \right]$) and using the fact that $\mathbb{E}\left[ S_u \right] = 1/\mu_u$ and $\mathbb{E}\left[ S_q \right] = 1/\mu_q$, we simplify Eq.~\eqref{avgSystemTimeOfUpdates} as
\begin{equation} \label{avgSystemTimeOfUpdates1}
\mathbb{E}\left[ {{T_{\rm{u}}}} \right] = {\rho _u}\mathbb{E}\left[ {{T_{\rm{u}}}} \right] + {\rho _q}\mathbb{E}\left[ {{T_q}} \right] + \frac{1}{\mu_u}.
\end{equation}

Similarly, the expected system time of the queries can be expressed as
\begin{equation} \label{avgSystemTimeOfQueries1} 
\mathbb{E}\left[ {{T_{\rm{q}}}} \right] = {\rho _u}\mathbb{E}\left[ {{T_{\rm{u}}}} \right] + {\rho _q}\mathbb{E}\left[ {{T_q}} \right] + \frac{1}{\mu_q}.
\end{equation}

Solving Eqs.~(\ref{avgSystemTimeOfUpdates1}) and (\ref{avgSystemTimeOfQueries1}) yields Eq.~\eqref{eq:fcfs_resptime} and the following:
\begin{equation} \label{avgSystemTimeOfUpdates2}
\mathbb{E} \left[ {{T_u}} \right] = \dfrac{{{\rho _q}/{\mu _q} + \left( {1 - {\rho _q}} \right)/{\mu _u}}}{{1 - {\rho _u} - {\rho _q}}}.
\end{equation}
Then, plugging Eq.~\eqref{avgSystemTimeOfUpdates2} into Eq.~\eqref{eq:paoi} and using the fact that $\mathbb{E}\left[ X_u \right] = 1/\lambda_u$, we obtain Eq.~\eqref{eq:fcfs_paoi} and complete the proof.
\end{IEEEproof}



The FCFS policy is a simple algorithm and is easy to implement in practice. However, a key limitation is that the FCFS policy does not provide a knob for prioritizing either updates or queries and thus cannot achieve a desired balance between service performance and information freshness. To that end, in the following subsections we will propose threshold-based policies that can prioritize either queries or updates and thus control the tradeoff between the response time and the PAoI.

\subsection{The Query-$k$ Policy}
In this subsection, we propose the Query-$k$ policy that sets a threshold $k$ for the query queue and prioritizes the queries whenever the length of the query queue reaches $k$. We will analyze the response time and the PAoI under this policy.

Specifically, the Query-$k$ policy functions in the following manner: (\romannumeral1) there is one single threshold $k$ for the query queue; (\romannumeral2) when the server is currently serving the update queue, the server has to switch from the update queue to the query queue instantly either if the number of queries reaches the threshold $k$ (thus preemption is allowed in this policy) or the update queue becomes empty; (\romannumeral3) no work of updates is lost due to the switches; (\romannumeral4) once the server switches to  the query queue, it needs to empty all queries waiting in the queue before it switches back to the update queue; (\romannumeral5) within each queue, FCFS is applied.

In the following, we will discuss three cases of the threshold value:
1) $k = 1$, 2) $1 < k < \infty $, and 3) $k = \infty$. 

\subsubsection{Threshold $k$=1} 
In this case, the server processes queries as long as the query queue is non-empty. Hence, the query queue always has a higher priority than the update queue. This model is equivalent to a preemptive priority queue with two classes of jobs \cite[Ch. 32]{harchol2013performance}.  

\begin{proposition} \label{pro:query-1}
Under the Query-$1$ policy, the expected response time is 
\begin{equation}
\mathbb{E}\left[ {{T_q}} \right] =  \dfrac{1}{{{\mu _q}}} + \dfrac{{{\rho _q}/{\mu _q}}}{{1 - {\rho _q}}},
\label{eq:query1_resptime}
\end{equation}
and the expected PAoI is 
\begin{equation}
\begin{array}{l}
\mathbb{E}\left[ {{A}} \right] 
 = \dfrac{1}{{{\lambda _u}}} + \dfrac{{1/{\mu _u}}}{{1 - {\rho _q}}} + \dfrac{{{\rho _q}/{\mu _q} + {\rho _u}/{\mu _u}}}{{\left( {1 - {\rho _q}} \right)\left( {1 - {\rho _q} - {\rho _u}} \right)}}.
\end{array}
\label{eq:query1_paoi}
\end{equation}
\end{proposition}

\begin{IEEEproof}
Consider a more general model where jobs are classified into $N$ classes~\cite[Ch. 32]{harchol2013performance}. Assume that Class 1 has the highest priority, Class 2 has the second highest priority, and so on. A job of higher class can preempt a job of lower class, but no work is lost due to preemption. Then, the expected system time for a job of Class $n$ (where $n=1,2,\dots$), denoted by $T(n)$, can be expressed as
\begin{equation}
{\mathbb{E}\left[ {T\left( n \right)} \right] = \dfrac{{\mathbb{E}\left[ {{S_n}} \right]}}{{1 - \sum\nolimits_{i = 1}^{n - 1} {{\rho _i}} }} + \dfrac{{\sum\nolimits_{i = 1}^n {{\rho _i}\dfrac{{\mathbb{E}\left[ {S_i^2} \right]}}{{2\mathbb{E}\left[ {{S_i}} \right]}}} }}{{\left( {1 - \sum\nolimits_{i = 1}^{n - 1} {{\rho _i}} } \right)\left( {1 - \sum\nolimits_{i = 1}^n {{\rho _i}} } \right)}}},
\label{eq:priority}
\end{equation}
where $S_i$ is the service time of jobs of Class $i$ and $\rho _i$ is the load due to jobs of Class $i$. 
In our model, there are two classes: the queries belong to Class 1, and the updates belong to Class 2. Hence, by plugging $n=1$ and $n=2$ into Eq.~\eqref{eq:priority} and using the fact that $\mathbb{E}\left[ {{T_q}} \right] = \mathbb{E}\left[ {T\left( 1 \right)} \right]$ and $\mathbb{E}\left[ {{T_u}} \right] = \mathbb{E}\left[ {T\left( 2 \right)} \right]$, respectively, we obtain Eq.~\eqref{eq:query1_resptime} and the following:
\begin{equation}
\mathbb{E}\left[ {{T_u}} \right] = 
\dfrac{{1/{\mu _u}}}{{1 - {\rho _q}}} + \dfrac{{{\rho _q}/{\mu _q} + {\rho _u}/{\mu _u}}}{{\left( {1 - {\rho _q}} \right)\left( {1 - {\rho _q} - {\rho _u}} \right)}}.
\label{eq:query1_Tu}
\end{equation}
Then, by plugging Eq.~\eqref{eq:query1_Tu} into Eq.~\eqref{eq:paoi} and using the fact that $\mathbb{E}\left[ X_u \right] = 1/\lambda_u$, we obtain Eq.~\eqref{eq:query1_paoi} and complete the proof.
\end{IEEEproof}

\subsubsection{Threshold $1 < k < \infty $}
Since the threshold $k$ is now larger than 1, the query queue has a higher priority than the update queue only when the threshold $k>1$ is reached. In other words, the query queue no longer has an absolute priority over the update queue. Hence, the analysis techniques used for the case of $k=1$ is not applicable here. Instead, we will exploit the techniques developed in \cite{boxma1995two} to analyze the response time and the PAoI. 

The work of \cite{boxma1995two} studies a similar two-queue model and consider a threshold-based policy. The authors
propose a method for calculating the probability-generating functions, which can be used for calculating the expected queue length. We will use the method of \cite{boxma1995two} in the analyses of the response time and the PAoI under the Query-$k$ policy.

We first define the following steady-state probabilities:
\begin{equation}
  \begin{split}
  {p_{ij}} &= \mathbb{P}\left( {{N_q} = i,{N_u} = j, Z = 1} \right),i \ge 1,j \ge 0,\\
  {q_{ij}} &= \mathbb{P}\left( {{N_q} = i,{N_u} = j, Z = 2} \right),0 \le i < k,j \ge 1,\\
  {r_{00}} &= \mathbb{P}\left( {{N_q} = 0,{N_u} = 0} \right),
  \end{split}
\end{equation}
where $Z=1$ if the server is serving the query queue and $Z=2$ if the server is serving the update queue. Apparently, the steady-state probabilities should satisfy the following:
  \begin{equation}
  {r_{00}} + \sum\limits_{i = 1}^\infty  {\sum\limits_{j = 0}^\infty  {{p_{ij}}} }  + \sum\limits_{i = 0}^{k - 1} {\sum\limits_{j = 1}^\infty  {{q_{ij}}} }  = 1.
  \label{eq:sumCon}
  \end{equation}
The corresponding probability-generating functions are
\begin{equation}
\begin{split}
  P\left( {x,y} \right) &= \sum\limits_{i = 1}^\infty  {\sum\limits_{j = 0}^\infty  {{p_{ij}}{x^{i - 1}}{y^j}} }, \\
 {Q_i}\left( y \right) &= \sum\limits_{j = 1}^\infty  {{q_{ij}}{y^{i - 1}}} ,0 \le i < k.
\end{split}
\label{eq:pgf}
\end{equation}
Using the probability-generating functions in Eq.~\eqref{eq:pgf}, we can express the expected queue length of the query queue as 
\begin{equation} \label{eq:query_k-Nq}
    \mathbb{E}\left[ {{N_{\rm{q}}}} \right] = {\left. {\frac{d}{{dx}}xP\left( {x,y=1} \right)} \right|_{x = 1}} + \sum\limits_{i = 0}^{k - 1} {i{Q_i}} \left( y=1 \right).
\end{equation}


We are now ready to state the main results for the Query-$k$ policy with $1 < k < \infty$.
\begin{proposition} \label{pro:query-norm-k}
Under the Query-$k$ policy with $1 < k < \infty$, the expected  response  time   is
\begin{equation} \label{eq:querk_k-Tq}
    \mathbb{E}\left[ {{T_q}} \right] = \mathbb{E}[{N_q}]/{\lambda _q},
\end{equation}
and the expected PAoI  is
\begin{equation} \label{eq:query_k-PAoI}
     \mathbb{E}\left[ {{A}} \right] = \dfrac{1}{{{\lambda _u}}} + \dfrac{{{\mu _{\rm{u}}}}}{{{\lambda _u}}} \cdot \left( {\dfrac{{{\lambda _{\rm{q}}}{\rm{/}}\mu _{\rm{q}}^2{\rm{ + }}{\lambda _{\rm{u}}}{\rm{/}}\mu _{\rm{u}}^2}}{{1 - \rho }} - \dfrac{{\mathbb{E}\left[ {{N_q}} \right]}}{{{\mu _{\rm{q}}}}}} \right),
\end{equation}
where $\mathbb{E}\left[ {{N_q}} \right]$ is given by Eq.~\eqref{eq:query_k-Nq}.
\end{proposition}

\begin{IEEEproof}
First, simply applying the Little's Law for the query queue (i.e., $\mathbb{E}[{N_q}] = \lambda_q \mathbb{E}\left[ {{T_q}} \right]$) yields Eq.~\eqref{eq:querk_k-Tq}. 

Next, by applying the Conservation Law \cite[pp. 236--238]{Gelenbe:2010:ASC:1830463}, we can also calculate the expected queue length of the update queue as
\begin{equation}
\mathbb{E}\left[ {{N_u}} \right] = \mu_u \cdot \left( {\dfrac{{{\lambda _{\rm{q}}}{\rm{/}}\mu _{\rm{q}}^2{\rm{ + }}{\lambda _{\rm{u}}}{\rm{/}}\mu _{\rm{u}}^2}}{{1 - \rho }} - \dfrac{{\mathbb{E}\left[ {{N_q}} \right]}}{{{\mu _{\rm{q}}}}}} \right).
\label{eq:query_k-Nu}
\end{equation}

Then, by applying the Little's Law for the update queue (i.e., $\mathbb{E}[{N_u}] = \lambda_u \mathbb{E}\left[ {{T_u}} \right]$) in Eq.~\eqref{eq:query_k-Nu} to get $\mathbb{E}\left[ {{T_u}} \right]$, plugging it into Eq.~\eqref{eq:paoi}, and using the fact that $\mathbb{E}\left[ X_u \right] = 1/\lambda_u$, we obtain Eq.~\eqref{eq:query_k-PAoI} and complete the proof.
\end{IEEEproof}

\emph{Remark:} To compute $\mathbb{E}\left[ {{N_{\rm{q}}}} \right]$ in Eqs.~\eqref{eq:querk_k-Tq} and \eqref{eq:query_k-PAoI}, we need to solve the global balance equations along with Eq.~\eqref{eq:sumCon}, express the probability-generating functions $P\left( {x,y} \right)$ and ${Q_i}\left( y \right)$'s in terms of $\lambda _q$, $\lambda _u$, $\mu _q$, and $\mu _u$, and plug them into Eq.~\eqref{eq:query_k-Nq}. The detailed derivation can be found in \cite{boxma1995two} and is omitted here.

\subsubsection{Threshold $k= \infty$} In this case, since the threshold of the query queue is infinity, the server switches to serving the queries only when the update queue becomes empty. Then, it keeps serving the queries. Only when the query queue becomes empty, the server switches to serving the updates. Therefore, the system reduces to the classical two-queue model with exhaustive service at both queues (i.e., all jobs waiting in the current queue will be served before the server turns to the other queue)~\cite{takacs1968two}. The work of \cite{takacs1968two} presents a method for deriving the distribution of waiting time for updates and queries. By using their method, we can obtain the expected system time for updates and queries, respectively, which can further be used to analyze the response time and the PAoI.

\subsection{The Update-$k$ Policy}
Similar to the Query-$k$ policy that priorities the queries, we propose another threshold-based policy, called the the Update-$k$ policy, which prioritizes the updates. 
Similarly, we will discuss three cases: 1) $k = 1$, 2) $1 < k < \infty $, and 3) $k = \infty$. 

\subsubsection{Threshold $k$=1}
In this case, the server always gives a higher priority to the update queue. Hence, the updates now belong to Class 1, and the queries belong to Class 2. We state the following proposition and omit the proof as it is similar to that of Proposition~\ref{pro:query-1}.

\begin{proposition} \label{pro:update-1}
Under the Update-$1$ policy, the expected response time is 
\begin{equation}
\mathbb{E}\left[ {{T_q}} \right] =\mathbb{E}\left[ {T\left( 2 \right)} \right] = \dfrac{{1/{\mu _q}}}{{1 - {\rho _u}}} + \dfrac{{{\rho _q}/{\mu _q} + {\rho _u}/{\mu _u}}}{{\left( {1 - {\rho _u}} \right)\left( {1 - {\rho _q} - {\rho _u}} \right)}},
\end{equation}
and the expected PAoI is 
\begin{equation}
\begin{array}{l}
\mathbb{E}\left[ {{A}} \right] = \mathbb{E}\left[ {{X_u}} \right] + \mathbb{E}\left[ {T\left( 1 \right)} \right]
= \dfrac{1}{{{\lambda _u}}} + \dfrac{1}{{{\mu _u}}} + \dfrac{{{\rho _u}/{\mu _u}}}{{1 - {\rho _u}}}.
\end{array}
\end{equation}
\end{proposition}


\begin{figure*}[tp]
	\begin{minipage}[t]{1\textwidth}
		\centering
		\subfigure[FCFS]{
			\label{fig:FCFS} 
			\includegraphics[width=0.32\textwidth]{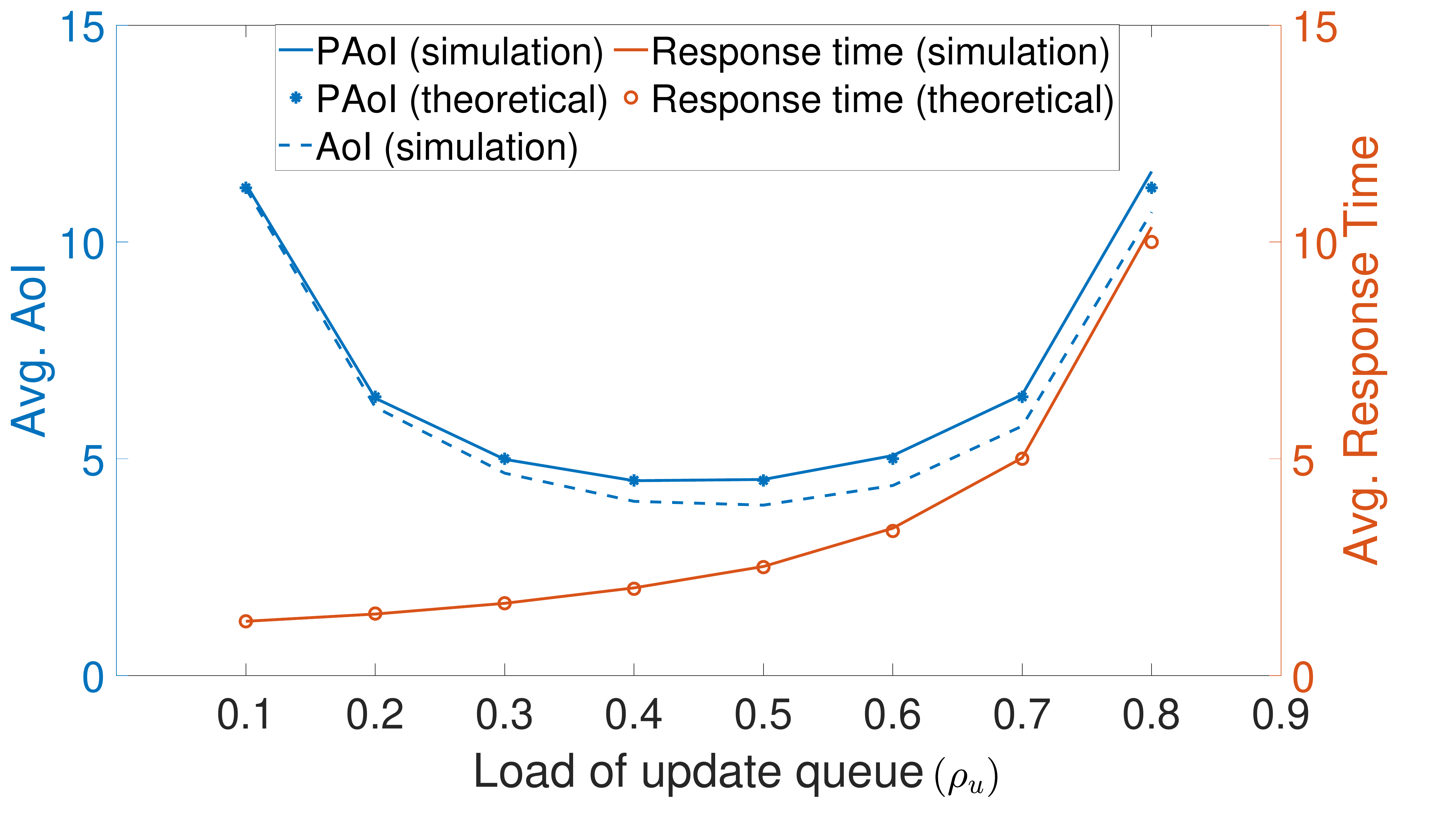}}
		\subfigure[Query-$1$]{
			\label{fig:Query-1} 
			\includegraphics[width=0.32\textwidth]{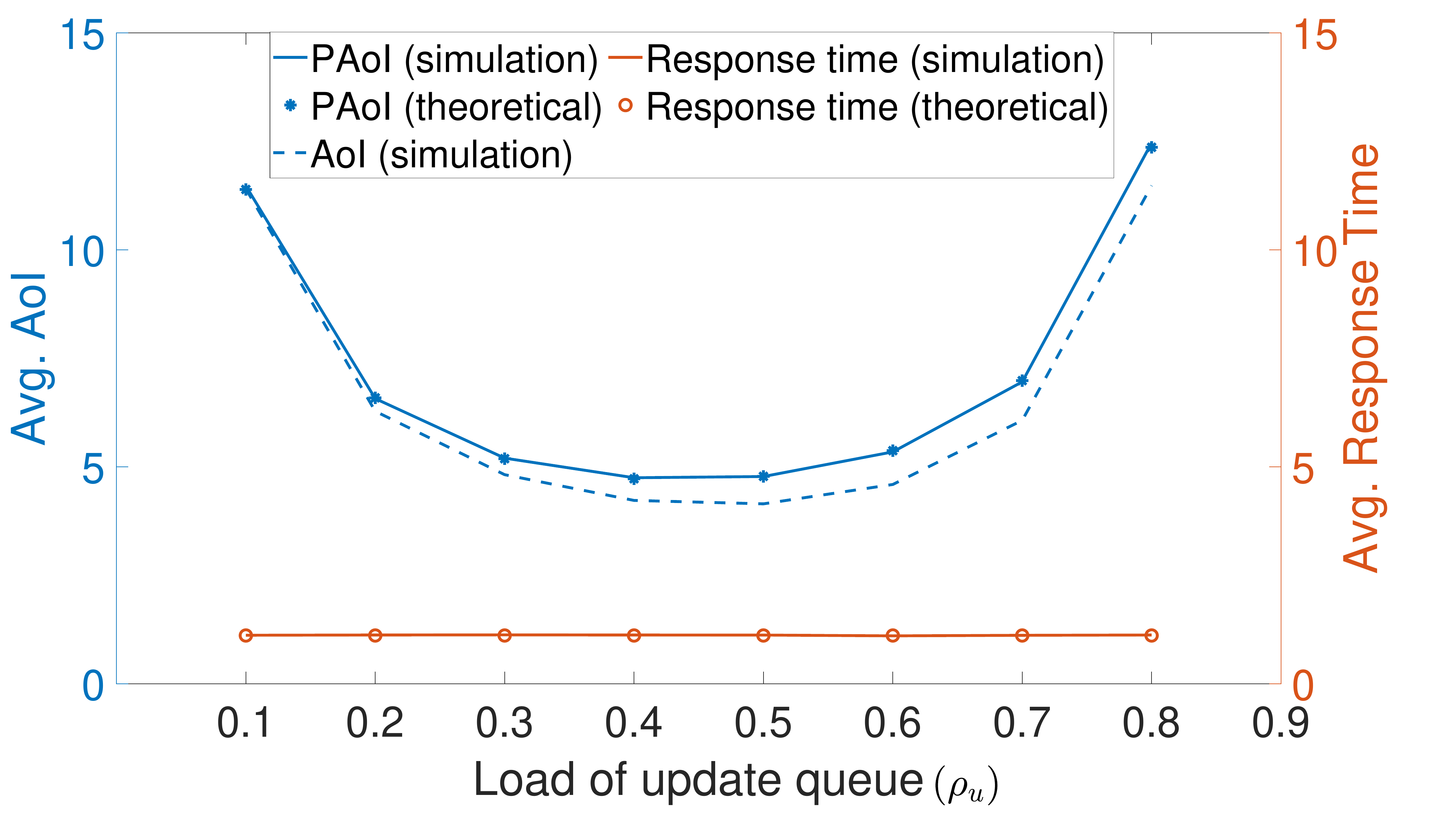}}
		\subfigure[Query-$3$]{
			\label{fig:Query-3} 
			\includegraphics[width=0.32\textwidth]{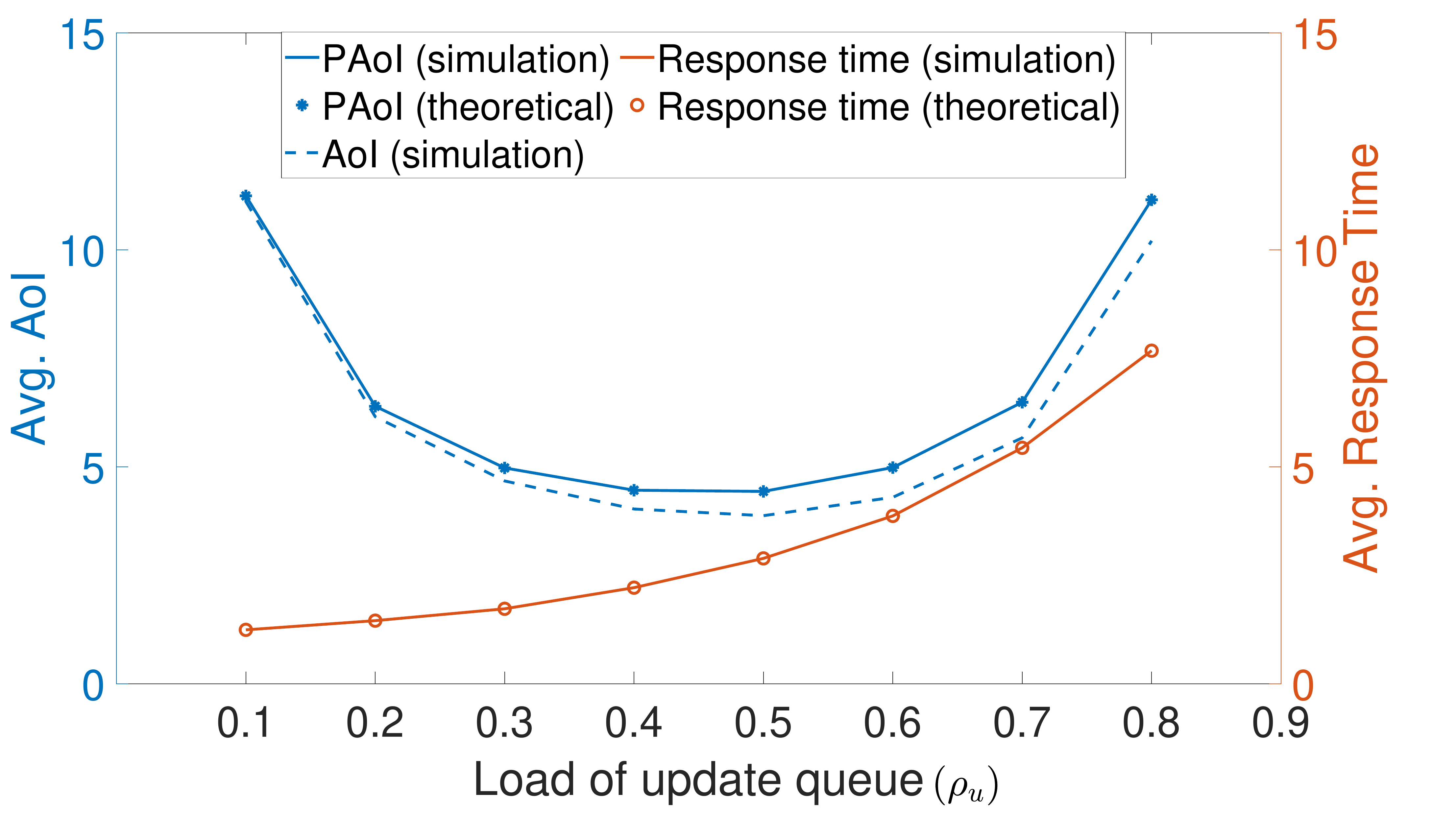}}
		\caption{Performance comparisons of different policies with varying update load (${\lambda _q} = 0.1$ and ${\mu _q} = {\mu _u} = 1$)}
		\label{fig:tradeoff-lambda} 
	\end{minipage}
\end{figure*}

\begin{figure*}[tp]
	\begin{minipage}[t]{1\textwidth}
		\centering
		\subfigure[FCFS]{
			\label{fig:FCFS_q} 
			\includegraphics[width=0.32\textwidth]{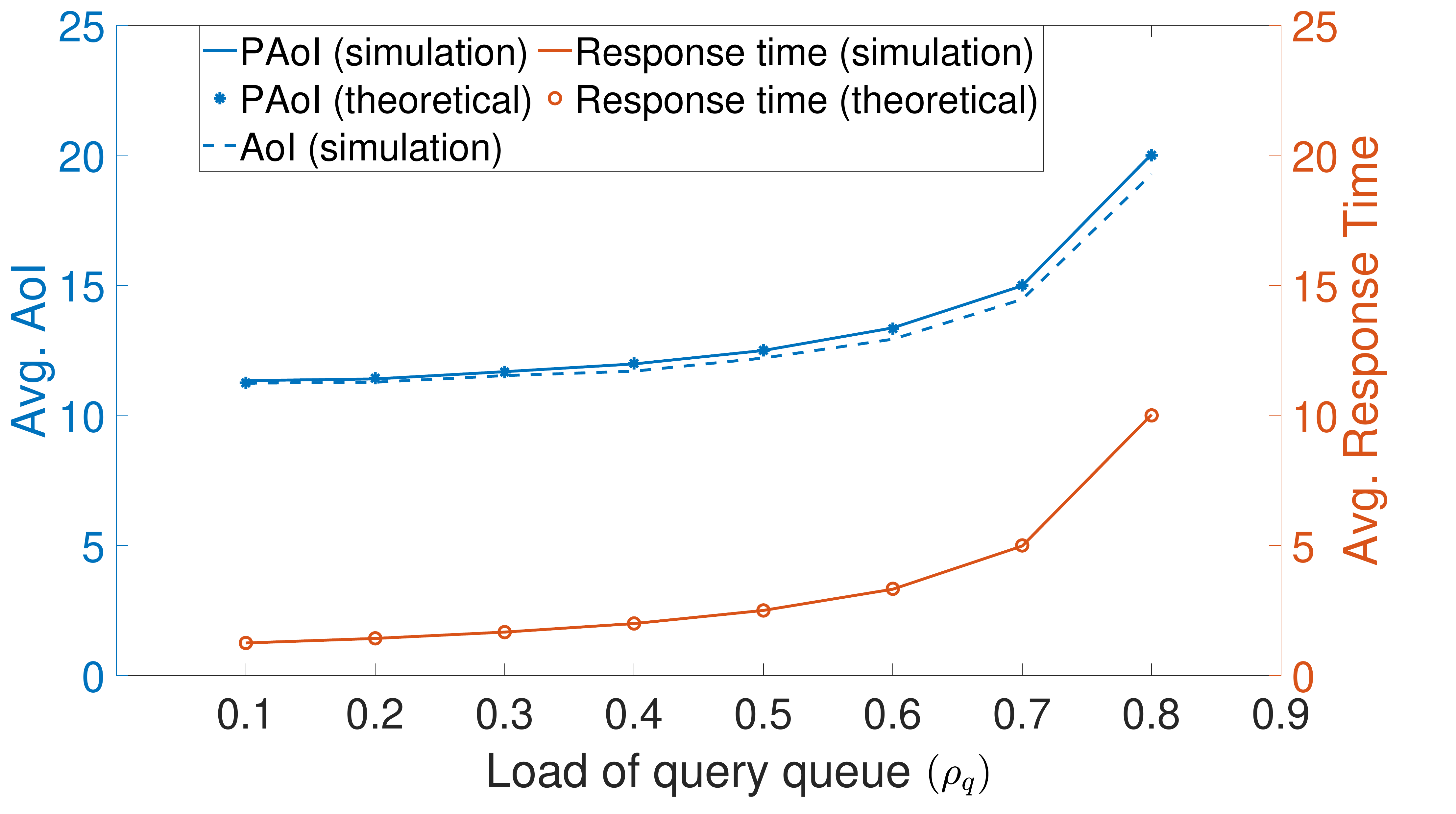}}
		\subfigure[Update-$1$]{
			\label{fig:Update-1} 
			\includegraphics[width=0.32\textwidth]{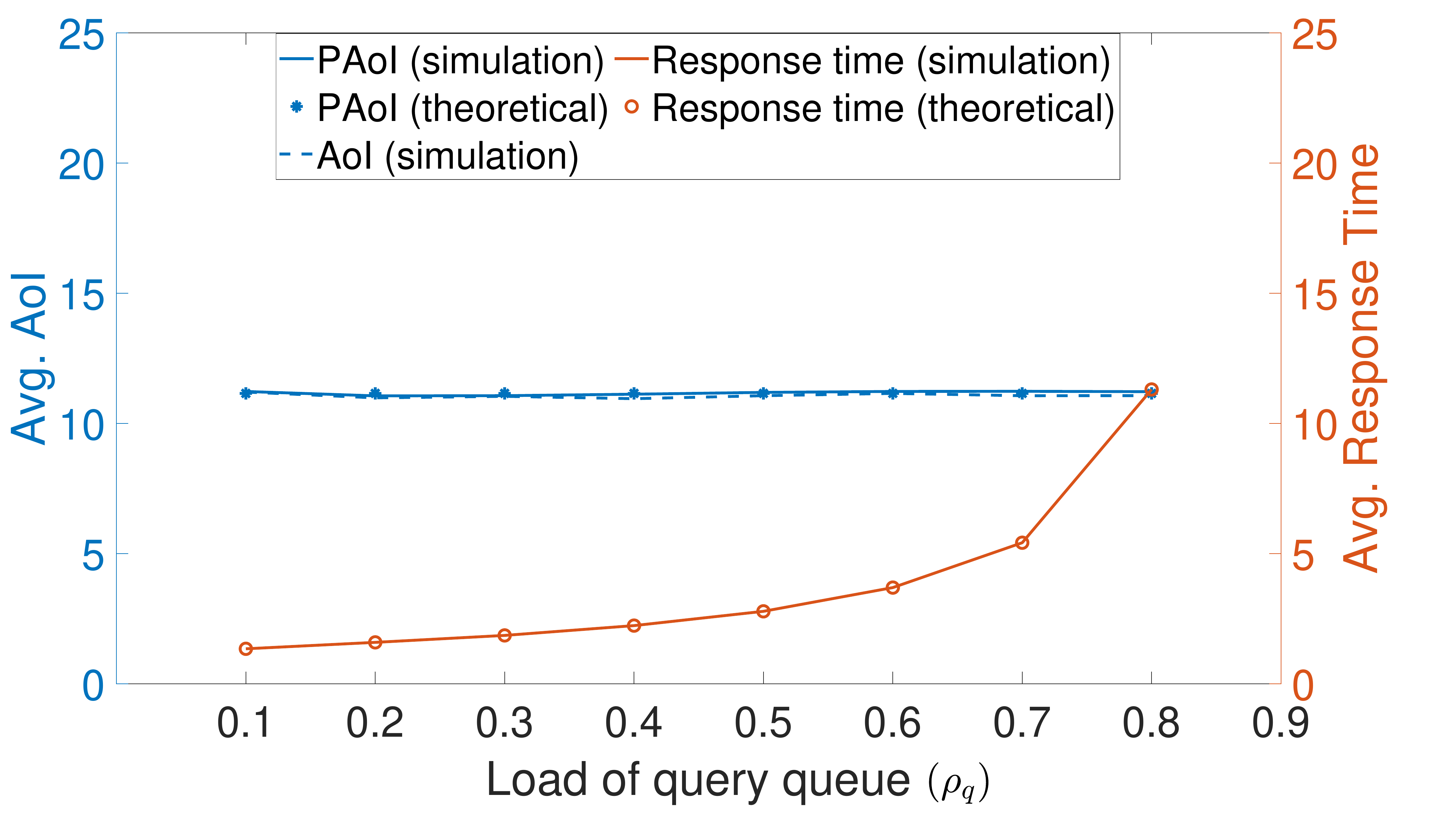}}
		\subfigure[Update-$3$]{
			\label{fig:Update-3} 
			\includegraphics[width=0.32\textwidth]{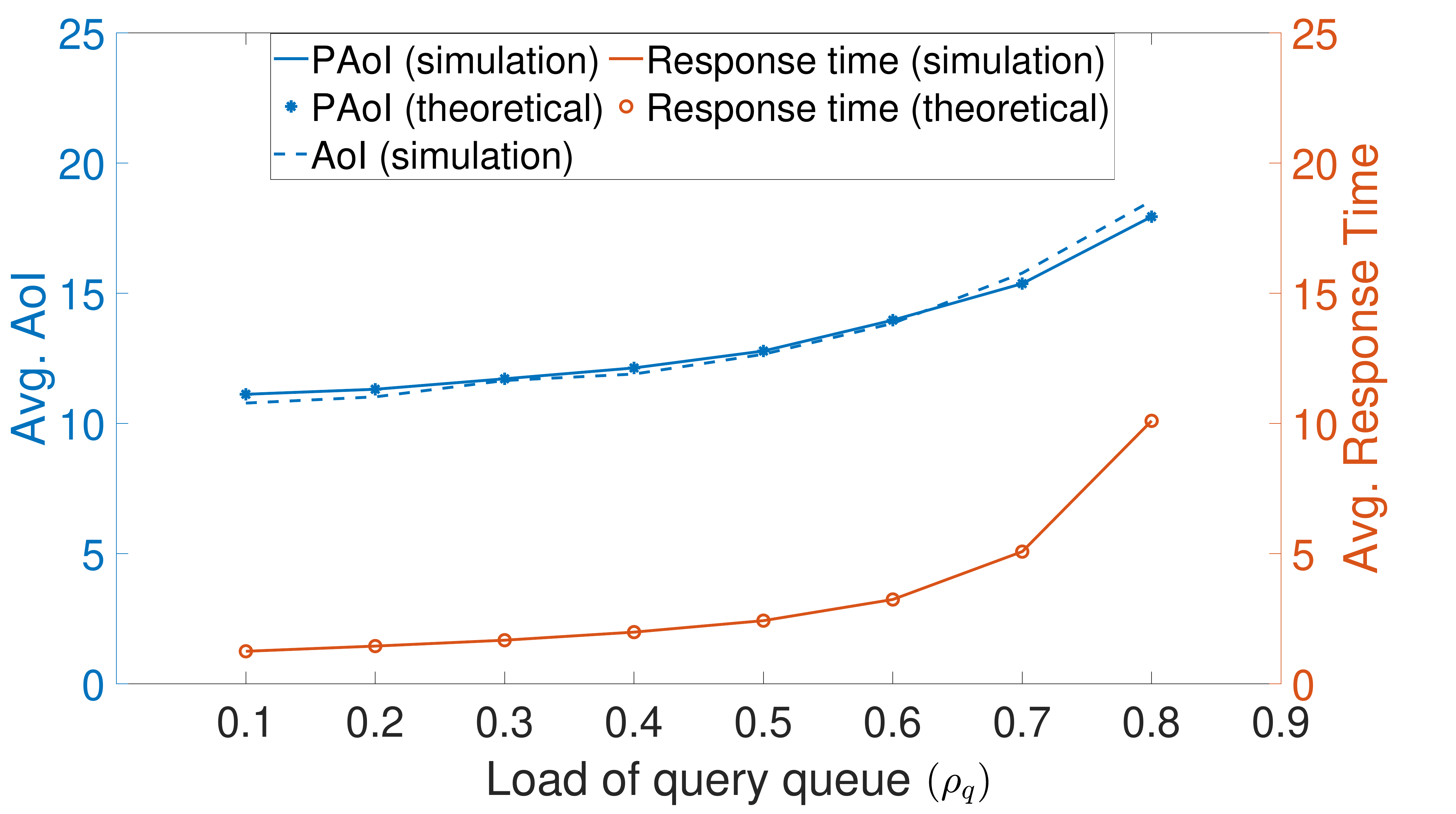}}
		\caption{Performance comparisons of different policies with varying query load (${\lambda _u} = 0.1$ and ${\mu _q} = {\mu _u} = 1$)}
		\label{fig:tradeoff-rou} 
	\end{minipage}
\end{figure*}

\subsubsection{Threshold $1 < k < \infty $}
This case is similar to the case of the Query-$k$ policy with $1 < k < \infty$. Following the same line of analysis as that in the proof of Proposition~\ref{pro:query-norm-k}, we can compute $\mathbb{E}\left[ {{N_u}} \right]$ using the techniques developed in \cite{boxma1995two} and analyze the expected response time and the PAoI. 
We state the main results in Proposition~\ref{pro:update-norm-k} and omit the detailed proof. 
\begin{proposition} \label{pro:update-norm-k}
Under the Update-$k$ policy with $1 < k < \infty$, the expected response time is
\begin{equation} \label{eq:update_k-Tq} 
\mathbb{E}\left[ {{T_q}} \right] =\dfrac{{{\mu _q}}}{{{\lambda _q}}} \cdot \left( {\dfrac{{{\lambda _{\rm{q}}}{\rm{/}}\mu _{\rm{q}}^2{\rm{ + }}{\lambda _{\rm{u}}}{\rm{/}}\mu _{\rm{u}}^2}}{{1 - \rho }} - \dfrac{{\mathbb{E}\left[ {{N_u}} \right]}}{{{\mu _u}}}} \right),
\end{equation}
and the expected PAoI  is
\begin{equation} \label{eq:update_k-PAoI}
    \mathbb{E}\left[ {{A}} \right] = 1/{\lambda _u} + \mathbb{E}\left[ {{N_u}} \right]/{\lambda _u}.
\end{equation}
\end{proposition}


\subsubsection{ Threshold $k=\infty$} Same as the Query-$k$ policy, the system reduces to the classical two-queue model with exhaustive service at both queues. The analysis will be exactly the same as that of the Query-$k$ policy with $k=\infty$.

	

\subsection{The Joint-$\left( {M,N} \right)$ Policy}
In the previous two subsections, we have been focused on threshold-based policies that prioritize either queries or updates. The analyses reveal the following insights: the priority is given to the queue with a threshold; the lower the threshold, the higher the degree of priority. Take the Query-$k$ policy for example. When $k=1$, the query queue always has a higher priority; when $k=\infty$, the query queue no longer has a higher priority, because the system reduces to the classical two-queue model with exhaustive service at both queues. Hence, one limitation of the single-threshold-based policies is that the priority is given to one queue only.

Next, we introduce the Joint-$\left( {M,N} \right)$ policy, where we jointly set thresholds $M$ and $N$ for the updates and the queries, respectively. This policy generalizes the Query-$k$ policy and the Update-$k$ policy and allows flexibly prioritizing updates or queries through choosing different values of $M$ and $N$.

Specifically, the Joint-$\left( {M,N} \right)$ policy functions in the following manner: (\romannumeral1) the update queue has a threshold $M$, and the query queue has a threshold $N$; (\romannumeral2) the server immediately switches to the queue whose queue length reaches its threshold and continues to serve this queue as long as the threshold of the other queue is not reached; (\romannumeral3) if both thresholds are reached, the server will serve the queue with a new arrival.


The Query-$k$ policy and the Update-$k$ policy are two special cases of the Joint-$\left( {M,N} \right)$ policy, where $(M=\infty, N=k)$ and $(M=k, N=\infty)$, respectively. When $1 < M < \infty$ and $1 < N < \infty$, the Joint-$\left( {M,N} \right)$ policy becomes more flexible in prioritizing updates and queries. We leave the analyses of the general Joint-$\left( {M,N} \right)$ policy as our future work. However, in Section~\ref{sec:simulation} we provide simulation results to demonstrate its advantages compared to the one-threshold-based policies.


 

\section{Numerical Results} \label{sec:simulation}

In this section, we conduct simulations to evaluate the response time and the PAoI of the proposed policies. We first consider the FCFS policy and demonstrate its limitations. Then, we show that the single-threshold-based policies (i.e., the Query-$k$ policy and the Update-$k$ policy) have the ability to effectively control the tradeoff between the response time and the PAoI. Finally, we demonstrate the flexibility of the Joint-$\left( {M,N} \right)$ policy. We implement and simulate these policies in Java. In the simulation results, each data point is the average of 10 runs, and each run lasts 20,000 time units. 
We also include our analytical results computed using Wolfram Mathematica for the purpose of comparison.

We first simulate the FCFS policy and assume ${\lambda _q} = 0.1$ and ${\mu _q} = {\mu _u} = 1$. The results are presented in Fig.~\ref{fig:FCFS}. The results show that both the average PAoI and the average AoI decrease first and then increase as the update load increases. When the update load is low, the PAoI and the AoI are large due to large inter-arrival times of the updates; when the update load is high, the PAoI and the AoI are also large due to large queueing delays. On the other hand, the response time keeps increasing as a larger update load can only worsen the congestion condition for queries.
Hence, when the update load is high, the response time and the PAoI can both be poor since the FCFS policy does not prioritize either queries or updates. 

Next, we consider the Query-$k$ policy and assume ${\lambda_q} = 0.1$ and ${\mu_q} = {\mu_u} = 1$. The results are presented in Figs.~\ref{fig:Query-1} and \ref{fig:Query-3}. We can observe from Fig.~\ref{fig:Query-1} that the average response time remains unchanged under the Query-1 policy since the queries are always given a higher priority than the updates, while the PAoI is only slightly larger than that under the FCFS policy (e.g., 12.46 vs. 11.62 when ${\rho_u} = 0.8$). 
Fig.~\ref{fig:Query-3} shows that under the Query-$3$ policy, the response time keeps increasing as the update load increases, but it is still better than that under the FCFS policy. Compared to the Query-$1$ policy, while the PAoI is a little smaller (e.g., 11.15 vs. 12.46 when ${\rho_u} = 0.8$), the response time becomes much worse. Therefore, one need to carefully choose the value of the threshold so as to effectively control the tradeoff between the response time and the PAoI. Note that the PAoI does not vary much under different policies due to a small query rate of $0.1$.

Similarly, we compare the FCFS policy with the Update-$k$ policy with different values of $k$, by assuming ${\lambda_u} = 0.1$ and ${\mu_q} = {\mu_u} = 1$ and varying the query load. The results are presented in Fig.~\ref{fig:tradeoff-rou}, where similar observations can be made. 

\begin{figure*}[tp]
	\begin{minipage}[b]{0.65\textwidth}
		\centering
		\subfigure[Query-$k$]{
			\label{fig:Query-k} 
			\includegraphics[width=0.45\textwidth]{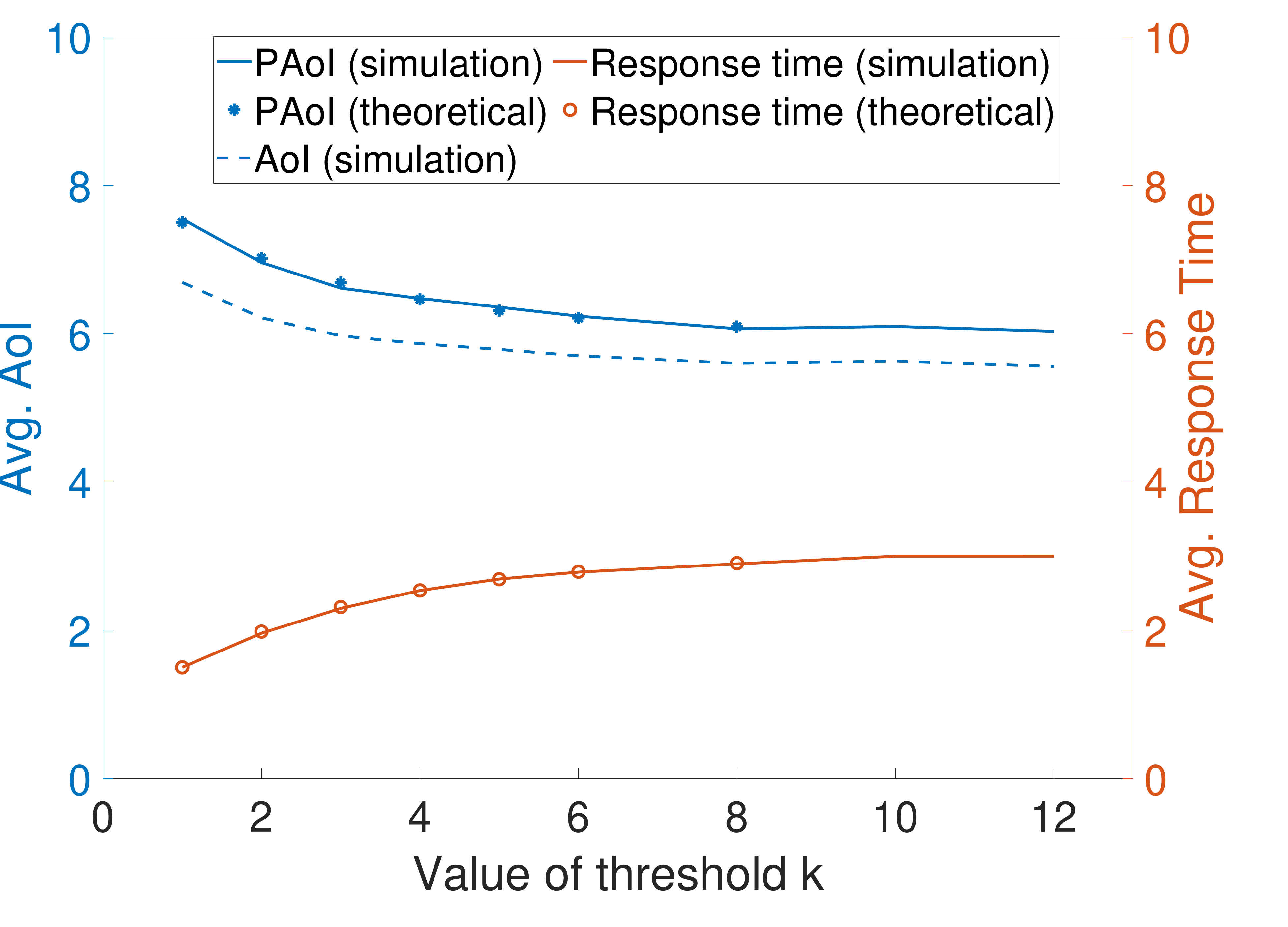}}
		\subfigure[Update-$k$]{
			\label{fig:Update-k} 
			\includegraphics[width=0.45\textwidth]{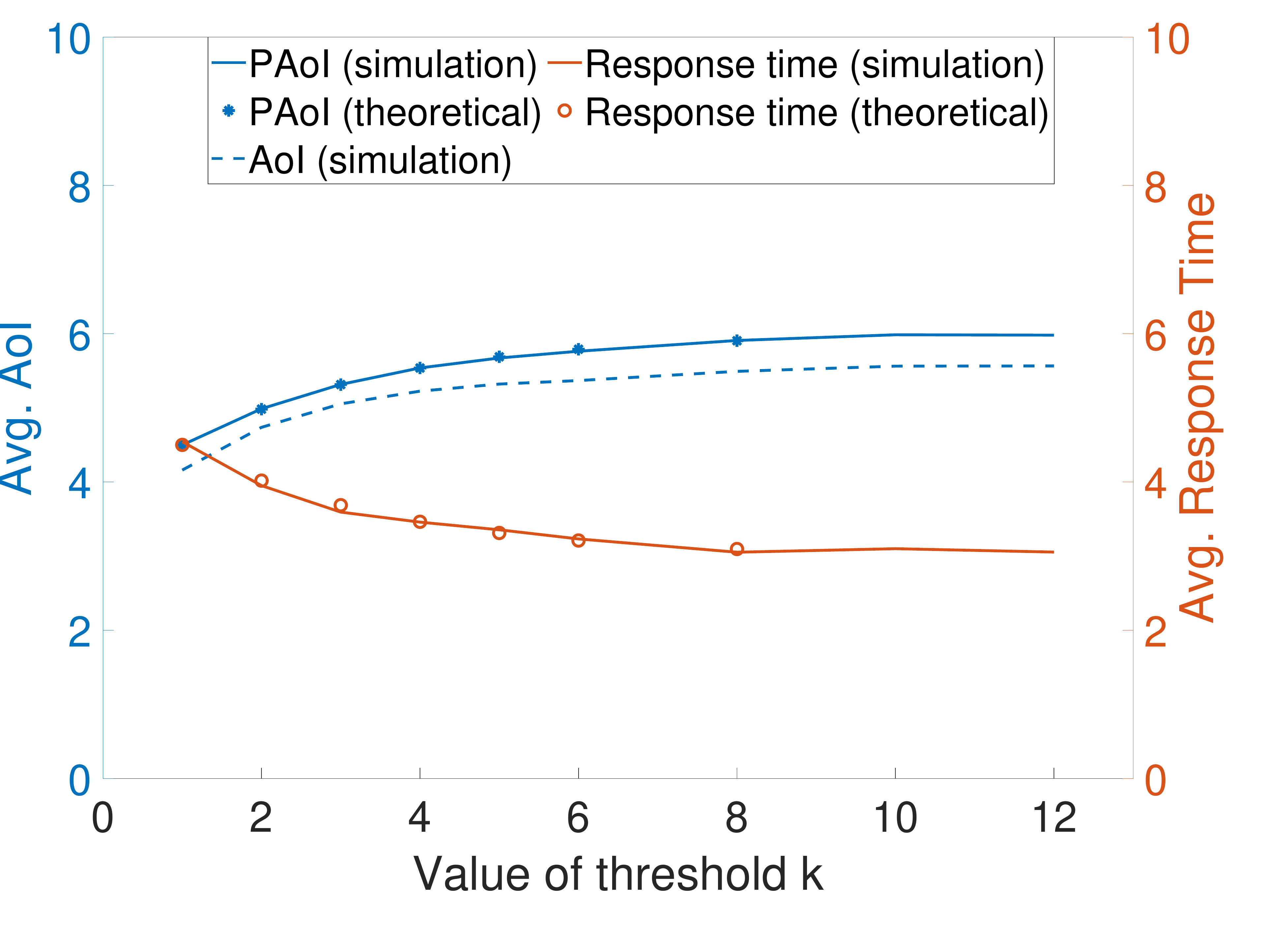}}
		\caption{Impact of the threshold on the single-threshold-based policies \\ (${\lambda_u}$ = ${\lambda_q}$ = $1/3$ and  ${\mu_u}$ = ${\mu_q}$ = $1$)}
		\label{fig:tradeoff-k} 
	\end{minipage}
	\begin{minipage}[b]{0.32\textwidth}
		\centering
		\includegraphics[width=0.9\textwidth]{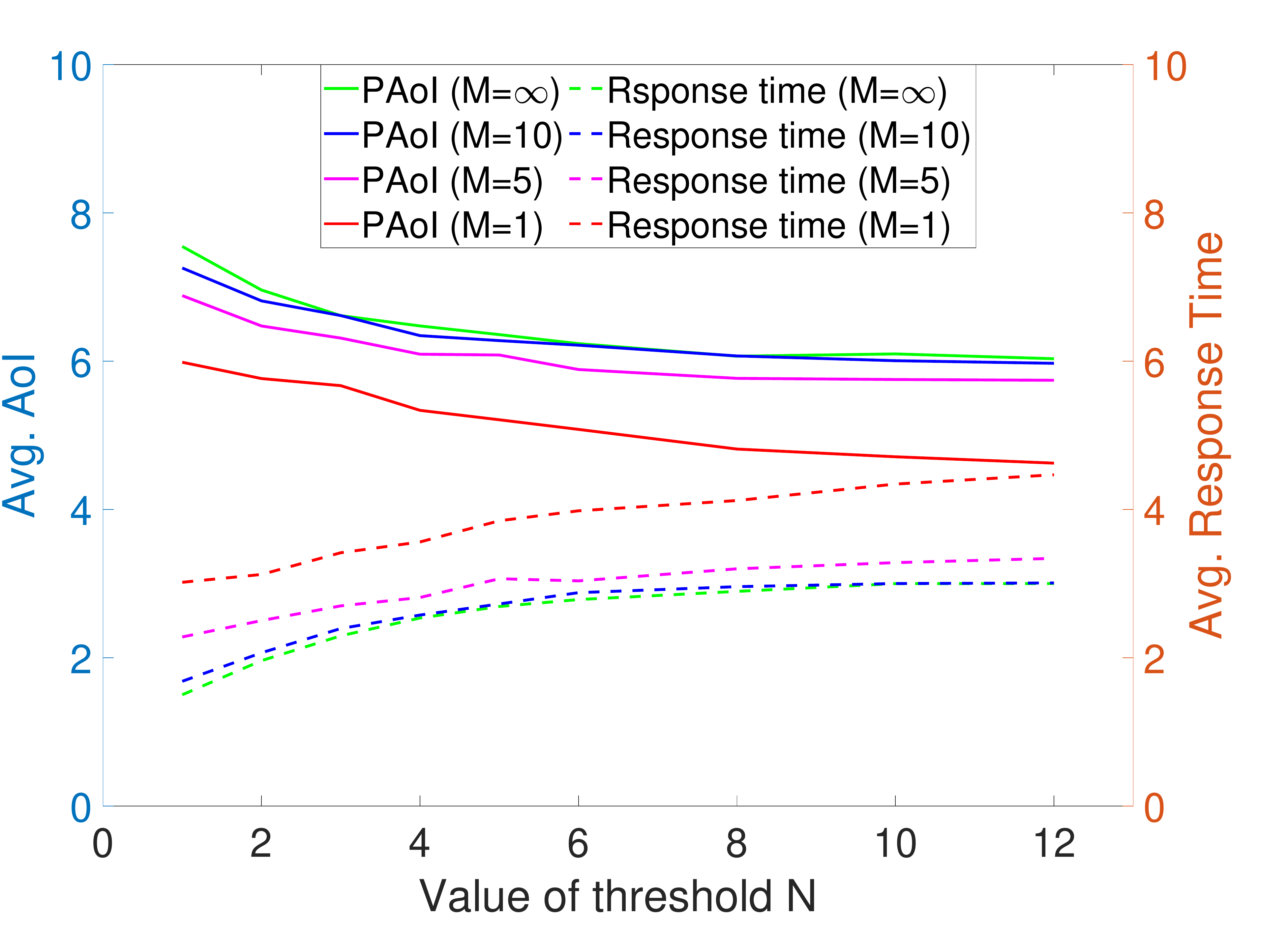}
		\caption{Impact of different values of the thresholds $M$ and $N$ on the Joint-$\left({M,N} \right)$ policy (${\lambda _u}$ = ${\lambda_q}$ = $1/3$ and  ${\mu_u}$ = ${\mu_q}$ = $1$)}
		\label{fig:Joint}
	\end{minipage}
\end{figure*}

Further, we investigate the impact of different values of the threshold under the threshold-based policies and present the results in Fig.~\ref{fig:tradeoff-k}. We assume  ${\lambda _u}$ = ${\lambda _q}$ = $1/3$ and  ${\mu _u}$ = ${\mu _q}$ = $1$. Fig.~\ref{fig:Query-k} shows that under the Query-$k$ policy, as the threshold $k$ increases, while the PAoI and the AoI decrease, the response time increases. This is because the degree of priority given to queries becomes lower as $k$ increases. Such behavior saturates when $k$ reaches a certain value (e.g., around $k=8$ in Fig.~\ref{fig:Query-k}). 
Similar observations can be made in Fig.~\ref{fig:Update-k}, which shows the results for the Update-$k$ policy.


Finally, we also simulate the Joint-$\left( {M,N} \right)$ policy. Assuming ${\lambda _u}$ = ${\lambda _q}$ = $1/3$ and  ${\mu _u}$ = ${\mu _q}$ = $1$, we investigate the impact of different values of the thresholds $M$ and $N$ on the response time and the PAoI. The results are presented in Fig.~\ref{fig:Joint}. We observe that the larger (resp., smaller) the value of $M$ (resp., $N$), the higher the PAoI and the lower the response time.
Therefore, the Joint-$\left( {M,N} \right)$ policy allows more flexibly prioritizing updates or queries through choosing different values of the two thresholds (i.e., $M$ and $N$).


\section{Conclusion} \label{sec:conclusion}
In this paper, we proposed a simple single-server two-queue model that captures the coupled scheduling between updates and queries for data-driven real-time applications. Aiming to address the natural tradeoff between service performance and information freshness in such applications, we proposed threshold-based scheduling policies that prioritize updates or queries and analyzed the response time and the PAoI in a rigorous manner. The simulation results further demonstrated that by properly choosing the values of the thresholds, the proposed policies can achieve the desired balance between service performance and information freshness.

Although this paper provides useful insights towards the tradeoff between the response time and the PAoI, there remain some open questions, which will be investigated in our future work. For example, it would be interesting to rigorously analyze the average AoI under the threshold-based policies and to systematically study the Joint-$\left( {M,N} \right)$ policy.  In addition, we have implicitly assumed that there was a negligible overhead for the server to switch back and forth between the query queue and the update queue. It would be interesting to investigate and characterize the impact of switching cost.

\bibliographystyle{IEEEtran}
\bibliography{reference}

\end{document}